\documentclass[apjl]{emulateapj}
\usepackage[figuresright]{rotating}
\usepackage{subfigure}
\usepackage{epic,eepic}
\usepackage{graphicx}
\usepackage{epstopdf}
\usepackage{CJK}
\usepackage[T1]{fontenc}

\shorttitle{sudden UV cutoff in a quasar}
\shortauthors{Guo et al.}

\begin{document}
\begin{CJK}{UTF8}{gbsn}

\title{The Optical Variability of SDSS Quasars from Multi-epoch Spectroscopy. III. A Sudden UV Cutoff in Quasar SDSS J2317+0005}

\author{Hengxiao Guo (郭恒潇) \altaffilmark{1,2,3}, Matthew A. Malkan\altaffilmark{2}, Minfeng Gu\altaffilmark{1}, Linlin Li\altaffilmark{1,3}, J. Xavier Prochaska\altaffilmark{4},  Jingzhe Ma\altaffilmark{5}, Bei You\altaffilmark{6}, Tayyaba Zafar\altaffilmark{7},Mai Liao\altaffilmark{1,3}}

\altaffiltext{1}{Key Laboratory for Research in Galaxies and Cosmology, Shanghai Astronomical Observatory, Chinese Academy of
Sciences, 80 Nandan Road, Shanghai 200030, China
hxguo@shao.ac.cn, hxguo@astro.ucla.edu}
\altaffiltext{2}{Department of Physics and Astronomy, UCLA, Los Angeles, CA, 90095-1547 USA}
\altaffiltext{3}{University of Chinese Academy of Sciences, 19A Yuquan Road, Beijing 100049, China}
\altaffiltext{4}{Department of Astronomy and Astrophysics, UCO/Lick Observatory, University of California, 1156 High Street, Santa Cruz, 95064, USA}
\altaffiltext{5}{Department of Astronomy, University of Florida, 211 Bryant Space Science Center, Gainesville, 32611, USA}
\altaffiltext{6}{Nicolaus Copernicus Astronomical Center, Polish Academy of Sciences, Bartycka 18, 00-716, Warsaw, Poland}
\altaffiltext{7}{European Southern Observatory, Karl-Schwarzschild-Strasse 2, 85748, Garching, Germany}

\begin{abstract}
We have collected near-infrared to X-ray data of 20 multi-epoch heavily reddened SDSS quasars to investigate the physical mechanism of reddening. Of these, J2317+0005 is found to be a UV cutoff quasar. Its continuum, which usually appears normal,  decreases by a factor 3.5 at 3000{\AA},  compared to its more typical bright state during an interval of 23 days.  During this sudden 
continuum cut-off, the broad emission line fluxes do not change, perhaps
due to the large size of the Broad Line Region (BLR), r  > 23 / (1+z)  days.   
The UV continuum may have suffered a dramatic drop out.
However, there are some difficulties with this explanation.
Another possibility is that  the intrinsic continuum did not change, but was temporarily blocked out, at least towards our line of sight. 
As indicated by X-ray observations, the continuum rapidly recovers after 42 days. 
A comparison of  the bright state and dim states would imply  an eclipse by a dusty cloud with a  reddening curve having a remarkably sharp rise shortward of 3500{\AA}. Under the assumption of being eclipsed by a Keplerian dusty cloud, we characterized the cloud size with our observations, however, which is a little smaller than the 3000\AA\ continuum-emitting size inferred from accretion disk models.  Therefore, we speculate this is due to a rapid outflow or inflow with a dusty cloud passing through our line-of-sight to the center.

\end{abstract}


\section{INTRODUCTION}
Extremely reddened quasars are an extraordinary sub-population that may help us understand quasar evolution \citep{hopkins2006},
and constrain the AGN unified model \citep{antonucci1993}. The UV and optical continuum of quasars, and possibly the X-rays, 
may be obscured by gas and dust during the early stage of black hole growth, until AGN feedback becomes
strong enough to clean out  the central region of the galaxy  \citep{dimatteo2005,hopkins2005,springel2005}. 
Alternately, in a somewhat different picture--the simplest version of AGN "unification"--
this absorption is produced by clouds in a hypothesized dusty torus.
If so, and the torus is always observed near its polar (empty) axis in Type 1 AGN, then absorption events would not be expected.

Another possibility is that  the absorption could occur many hundreds of parsecs further out \citep{mgt98},
in which case only slow long-term changes in absorption would be possible.
In a more complicated version of the unification model, the torus is clumpy.
Then an observer might see some circumnuclear clouds moving into and out of the line-of-sight.  
If the absorbers were large enough, they could temporarily convert  a pure Type 1 AGN into  a Type 1.8 or 1.9 
by eclipsing the broad components of H${\beta}$ \citep{osterbrock1981}, making a so-called "changing-look" AGN.
To date only a small number of changing-look Seyfert 1s have been found,  
and they all show the nonstellar
continuum emission weakening dramatically when the broad line components fade or even disappear
\citep{collin1973,tohline1976,lyutyj1984,cohen1986,storchi1993,aretxaga1999,eracleous2001,denney2014,sanmartim2014,barth2015}. 
Even factor of 2 changes in broad H${\beta}$ are rare in Seyfert 1's, having been seen in only about 4\%\ of the pairs of repeated observations in
a 6-year monitoring campaign of 13 AGN \citep{rosen92}. The timescale for these changes ranged from 30 to 100 days.

Recently \cite{lamassa2015} found the  changing-look quasar SDSS J015957.64+003310.5 at a redshift  z = 0.31, 
which transitioned from Type 1 to Type 1.9. 
Subsequently, several more changing-look quasars were uncovered from repeated Sloan Digital Sky Survey (SDSS) observations 
\citep{ruan2015,runnoe2015}. Now many researchers \citep[e.g.,][]{macleod16} are carefully examining all the SDSS galaxies and quasars systematically,
or also searching the Time Domain Spectroscopies Survey (TDSS) \citep{morganson15,ruan2016}, to find more changing-look AGN. 

The physical mechanism of this changing-look behavior is still an open question.  Changes in
continuum reddening could be caused by the transverse motion of an intervening dust cloud \citep{goodrich1989,mgt98,risaliti2009,wang2012}. 
The odds of such an alignment leading to a rapid eclipse would be enhanced for clouds relatively near the galactic center.
However, \cite{lamassa2015} suggest that their observed changing-look behavior--a correlated dimming of both continuum {\it and} the 
broad Balmer lines--
does not match their models of what such a dusty cloud should produce. 
Furthermore, the eclipses in most quasars occur over a timescale $\sim$ 10 years, which is a surprisingly short time for clouds to cross 
the entire broad line region at typical galactic velocities. 
Therefore, they consider an alternative in which  luminosity changes occur in the accretion disk, where the variability timescales should  be much
shorter \citep{lamassa2015,ruan2015,runnoe2015}.

In this paper, we report the first discovery of a UV cutoff quasar with redshift z = 0.32. 
This source SDSS J231742.60+000535.1 (hereafter J2317+0005) was initially observed in DR1 and also included in Stripe 82 survey. 
J2317+0005 was observed three times on 2000 Sep 29, 2001 Sep  25 and Oct 18, respectively. 
What is extraordinary is that  it transitioned to a dim state, and then recovered 
in a {\it few weeks}. We re-observed it on 2015 Sep 18 with the Kast Double Spectrograph on the Shane 3-meter telescope,  
finding it still back in the bright state. 

Our paper is organized as follows. In Section 2, we describe our sample selection and datasets. 
In Section 3, we will give results and discuss the possible origins of heavily reddened changing behavior and constrains the properties of clumps. 
Throughout this paper, a cosmology with $H_{0}=70\rm ~km~ s^{-1}~Mpc^{-1}$, $\Omega_{\rm m}=0.3$ and $\Omega_{\Lambda}=0.7$ is adopted.

\section{Quasar Sample and Data}
With the aim of investigating  color variability in quasars, we collected 2169 multi-epoch quasar spectra with 
signal to noise ratio S/N $\ge$ 10 and variability larger than 10$\%$ from the SDSS Data Releases 7 $\&$ 9.  
Out of this large sample, 20 heavily reddened quasars were carefully selected  when visual inspection revealed  their 
peculiar spectra 
(see the details, \cite{guo16}).  
Here we highlight one source, J2317+0005, which showed the most extreme reddening and most rapid changing behavior in just 65 days. 

J2317+0005 is a variable source in the very well-observed SDSS Stripe 82 \citep{sesar2007}.
This object has 54 epochs of photometric data in {\it u,g,r,i} and  {\it z} bands from 1999 Oct 16 to 2007 Oct 29 (see, Figure 2) \citep{york2000}. 
This quasar is also detected by the Two Micron All Sky Survey (2MASS) in the {\it J}, {\it H}, $\it {K_{s}}$ bands \citep{skrutskie2006}, 
the  United Kingdom InfraRed Telescope (UKIRT) Infrared Deep Sky Survey (UKIDSS) in the {\it Y, J, H, $K_{s}$} bands \citep{lawrence2007}, 
together with the W1, W2, W3, W4 bands in  the Wide-Field Infrared Survey (WISE) \citep{wright2010}. 
It was also imaged by $\it GALEX$ in  the FUV and NUV windows \citep{morrissey2007}. 
All observed photometric data are list in Table \ref{table1},  uncorrected for extinction.

Three SDSS spectra are found in the DR 7 quasar catalog \citep{shen2011}.  
The spectral wavelength coverage is 3800 - 9200 {\AA}, with resolution R $\sim$ 2000. 
In addition, low resolution spectra are retrieved from {\it GALEX}, covering the wavelength ranges from 1344 to 1786{\AA} and 1771 to 2831{\AA}. 
In order to check the current brightness of  this quasar, we observed J2317+0005 again using the Shane 3-meter telescope,
equipped with the Kast Dual Spectrograph at Lick observatory.  
Our measurements from the blue arm of the spectrograph, using a 452 lines $mm^{-1}$ grism, cover 3150 - 6200{\AA}  at a scale of 1.41 {\AA} $\rm pixel^{-1}$. 
Meanwhile, a 1200 lines $mm^{-1}$ grating was utilized in the red arm at a dispersion of  1.17 {\AA} $\rm pixel^{-1}$, covering  5800 to 7310{\AA}. 
A 2'' slit was used for our 1800s exposure (see Figure 1 and Table 2). 
Calibration frames including arc lamps and dome flats were observed each afternoon, and flux standards were observed during twilight. 
Data reduction was performed following the Standard IRAF \footnote[1]{\tt http://iraf.noao.edu/} routines. 

The $\it XMM-Newton$ archival data for J2317+0005 is serendipitous, i.e.,  
this quasar appeared in the field of the view of observations which targeted another scientific object (NGC 7589). 
Two observations on 2001 Jun 3 (ObsID: 0066950301) and Nov 28 (ObsID:0066950401)  were reduced with standard SAS threads: 
filtering for flaring, creating good time intervals and extracting the spectra from all three instruments (PN, MOS1, MOS2). 
With 12ks and 13ks exposure time in full-window mode,  251 and 420 photons were detected in the two MOS1 observations
(the best X-ray data for J2317+0005), respectively. 
To remove any artifacts from the final spectra,  we use the same parameters, such as source and background radius, 
on the two sets of raw data. Next, the spectra are fitted by an absorbed power-law model,
with one absorption component fixed to the Galactic value (3.74 $\times$ $10^{20} ~\rm cm^{-2}$), 
while another absorption is a free parameter at the redshift of the quasar. 
Figure 3 shows the absorbed power-law fits to the  two observations. 
Neither observation shows significant absorption above the Galactic value.

In Figure 2, we plot five-band magnitudes over the multiple SDSS observations taken during eight years in Stripe 82. 
Unfortunately, there is a gap in photometric coverage around 2001 Oct 18 (MJD = 52200).
So we also calculate the five-band magnitudes from the spectra to fill in the light curve. 
However, we find the {\it u} band point at 52200 is much fainter than any other epochs during the entire 8 years.  
Due to the strange shape of the green spectrum (MJD = 52200), we have searched available information to confirm that it is a correct spectrum, without observational problems. Though its plate (plateID = 680) flag is marked as "MARGINAL", 
it is still acceptable. Flags of ${\rm SN\_median}$ = 19 and {\rm Zwaring} = 0 indicate good resolution and no warnings for this spectrum. 
The inverse variance of the green one is even smaller 
than other epochs. The APBIAS flag, with which sources are used to test for an offset fiber effect, is not marked for this spectrum. 

Furthermore, we carefully checked the raw data of this spectrum. 
The observational log \footnote[2]{\tt http://data.sdss3.org/sas/dr12/sdss/spectro/redux/26/0680/} shows this object 
was not included in the unmatched fiber list in that plate. 
The most important confirmation is three independent blue spectra consistent with each other (see Figure 4), 
which make up the blue part of green epoch together,  and the calibration processes were confirmed by ourselves with the raw calibration 
files \footnote[3]{\tt http://data.sdss3.org/sas/dr12/sdss/spectro/data/52200/}.   

Images are also checked to ensure there is no bright nearby object contaminating it. 
Then, we coadd all of 54 r-band images to get a 3-magnitude deeper quasar image, and find it can be fitted by a 14.38 mag PSF 
and a 17.63 mag Sersic profile perfectly with GALFIT \citep{peng2002}. 
That indicates the contamination of the host galaxy is only $\sim$ 5$\%$, as expected for such
a luminous quasar.

\section{RESULTS and Discussions}
\subsection{Rapid Variability}
Quasar J2317+0005 is an unresolved almost purely point-like object, so SDSS Point Spread Function (PSF) magnitudes are adopted in our paper.  
In Figure 1, both spectra and photometry are corrected for galactic extinction (E(B-V) = 0.044)\citep{schlegel1998}, 
assuming a standard reddening law \citep{cardelli1989} at zero redshift.  
All the observations of J2317+0005 are broadly consistent with the quasar composite spectrum (gray line), 
which is derived from the UV/optical composite of $\sim$1800 SDSS quasars \citep[$\lambda$ $< $7000{\AA};][]{vanden2001} 
and the NIR composite of 27 quasars \citep[$\lambda$ $\ge$ 7000{\AA};][]{glikman2006}, except for one strange epoch (green line).  
The continuum of this spectrum shows an obvious UV-deficiency at $\lambda$ $<$ 3500 {\AA} without any absorption lines. 
The continuum flux at 3000{\AA} is about 3.5 times smaller than at the other epochs. 

Following the specific analysis in \cite{guo2014}, 
we find that the total flux of emission lines are consistent with no variability, even though the UV continuum varies a lot (see Table 3). 
( The consistency of emission line fluxes further indicates that there was no flux calibration error in the spectra, since that would have
brought down both lines and continuum equally. )
The most impressive thing is  that this UV drop-off lasted  only 23 days. 
The time might be so short that the BLR, which is far larger than the UV-continuum-emitting region,
likely did not have sufficient time to respond to the UV continuum change. 

This rapid major  change  is rare in the optical,  but common in X-ray absorptions \citep{turner2008,bianchi2009}. 
In X-rays, discrete warm/cold absorbers can change on timescales of  hours to months.  
So, this heavily reddened spectrum could be related to X-ray absorption events. 
In Figure 3, an X-ray spectrum observed 42 day later by XMM-Newton shows no evidence of absorption.
This implies to us that at least by that time, the far-UV flux had probably recovered to its normal bright level, though there are no measurements of it directly.

The rapid decrease in the UV continuum of our quasar, while its emission line fluxes remained constant,
could in principle be produced by either mechanism described in the Introduction--intrinsic (disk instability) or extrinsic (eclipse). 
Either scenario would require special circumstances, as we shall now examine.
In either case, there may have been a dramatic change in the illumination falling on the 
broad emission line region.
If so, then its lack of variability sets a lower limit on the size of the BLR about 17.4 light days.
Compared with the BLR lags detected in other quasars \citep{shen16}
this size limit is reasonable for the observed $\rm H{\beta}$ luminosity of 3.6 $\times 10^{42}~\rm erg~s^{-1}$

\subsection{Intrinsic Variation in the Inner Accretion Disk}

Since this source is included in the DR7 quasar catalog \citep{shen2011}, it
has an estimated approximate black hole mass, based on the width of its broad hydrogen lines
( $M_{\rm BH} = 10^{8.42 \pm 0.22}$$\rm M_{\rm \bigodot}$), and Eddington ratio ($L_{\rm bol}/L_{\rm Edd}$= 0.11).  
Assuming an average spectral energy distribution, 
they use its optical luminosity to estimate the bolometric luminosity $L_{\rm bol} = 10^{45.56 \pm 0.33}$ $\rm erg~s^{-1}$.

If the UV continuum--which dominates the bolometric luminosity of this powerful
quasar--is optically thick thermal emission as proposed by \citep{malkan1982} and others, 
then it must come from  a relatively large surface area.  The maximum possible emissivity
is given by the Planck function, which at 3000\AA\ for gas at $T_{\rm eff}$ = 27,000 K
requires a surface area of $10^{31}$ $\rm cm^2$.
This simple estimate ignores the actual complexities of an optically thick accretion disk.
On the one hand, some 3000\AA\ light is emitted by inner regions of the disk at even higher
temperatures, and their higher emissivity can be further boosted by relativistic Doppler shifts
\citep{malkan1983}.  But on the other hand, this will be decreased by gravitational redshifts
in the inner disk regions.  Fortunately, many models of optically thick accretion disks have
been analyzed for other quasars similar to J2317+0005.
In particular, the detailed disk fit to the quasar PKS 0405-123 shows that half of its 3000\AA\  
continuum arises at radii smaller than 50 Schwarzschild radii \citep{malkan1983}. The accreting black hole
in PKS 0405-123 is estimated to have a similar size.

Since PKS 0405-123 is about an order of magnitude more luminous than J2317+0005, 
we can bracket our estimate with another accretion disk model fit to a bright Seyfert 1
nucleus which is also believed to have an accreting black hole of almost comparable mass,
in NGC 5548. It is an order or magnitude less luminous than J2317+0005. 
A very similar detailed accretion disk model fit for the central engine in NGC 5548 predicts that 
80\% of its 3000\AA\ continuum arises from radii smaller than 12 Schwarzschild radii
\citep{krolik1991}.
The Eddington luminosity ratio for J2317+0005 appears to be equally bracketed
by these two well-studied AGNs. Thus  we will assume that the UV-emitting region
in J2317+0005--which was effectively turned off for ~ a month--has a radius of about
30 Schwarzschild radii.
For the assumed $M_{\rm BH}$,  this corresponds to $\sim$ 2 $\times 10^{15}$ cm, with an uncertainty of a factor of two. 
At this radius, the viscous accretion timescale is far longer than the observed variability
\cite{lamassa2015}.
However, the orbital timescale at this radius is about one month, and the smallest detectable
variations in AGN luminosity are thought to occur on the disk orbital timescale 
\citep{webb00,edelson14}. 
Thus some extraordinarily violent disk instability could-- {\it in principle}--shut down and restart
the thermally produced UV luminosity on a month timescale.
But we emphasize that the observed amplitudes of variations on such a short timescale
are normally only ~5\% \citep{webb00}.
An event with an amplitude of  80\%, as we observe in J2317+0005, 
would be virtually unprecedented among thousands of observed quasar light curves,
excluding a handful of blazars, where the emission is produced nonthermally
in a relativistic jet. 
Nonetheless, when trying to explain an extremely rare variability event, we cannot rule out
this extraordinary possibility.

\subsection{Absorbing Cloud Properties}
 
The other possible explanation for the UV turnoff is that it was a temporary eclipse of the UV continuum by
a dusty absorbing cloud moving across our  line-of-sight. 

As we shall see, the absorbing cloud cannot reside in the host galaxy hundreds of parsecs from the center,
That scenario for absorbed AGN was explicitly proposed as the "Galactic Dust Model" of \cite{malkan1998}.  
But the absorption variability time-scales  
would be one or two orders of magnitude too long.  
We are therefore lead to consider  their other scenario, where the dust cloud is a clump in an "Accreting Torus", much closer
to the central engine  \citep{jiang2013,zafar2015}. 

Recent IR observations at high spatial resolution suggest that the torus size might be no more than a few pc \citep{jaffe2004}.
In this region the gravitational potential is dominated by the mass of the central black hole \citep{nenkova2002,nenkova2008a,nenkova2008b,elitzur2006}.
Here we consider the possibility that a dusty clump, which might be associated with such a dust torus,
temporarily orbited in front of our line-of-sight, which is normally unblocked.
Our observations provide several independent constraints on what kind of optically thick
cloud could be consistent with such a scenario.
We illustrate these on a plot of cloud velocity transverse to our line-of-sight to the quasar versus cloud diameter, in Figure 5.
Assuming the cloud follows Keplerian motion, its orbital velocity is   $v_{orb }$= 1000 $\rm km~s^{-1}$ / Sqrt  $R_{\rm pc}$,
where $R_{\rm pc}$ is the distance to the central black hole in pc, and we have adopted the black hole mass estimate above.
(Note that if some of this velocity is not tangential, rather than purely transverse to the black hole in the nucleus,
then R would need to be even smaller than this estimate, giving an even stricter constraint.)

As described above, 
for the black hole in J2317+0005, 30 Schwarzschild radii correspond to a disk {\it diameter}
of 4 $\times$ $10^{15}$ cm.  To eclipse up to 80\%\ of this region would require an absorbing
cloud even larger than this lower limit, which is shown by the the horizontal lower limit in Figure 5.
Given the various uncertainties, we consider this estimate uncertain by $\pm$ 50\%, shown
as a horizontal grey band in Figure 5.

The 65-day limit on the eclipse duration derived from the optical and X-ray spectra constrain the absorber to have:
a) relatively small size and/or b) high velocity: t = Size/Velocity < 65 light days.  This upper limit is shown by the 
black diagonal line in Figure 5.
However these parameters are difficult to reconcile with the additional requirements that:
c) the cloud be large enough to almost completely cover the UV-continuum-emitting region; while at the same time
d) the dust grains were  not evaporated by the quasar continuum radiation.

At small distances R from the central engine, the dust cloud should be exposed to an intense quasar radiation field.
In the absence of shielding, the dust grains should come to an equilibrium temperature which we will
assume is below the dust sublimation temperature, believed to be between 1500 and 2000 K.
The continued survival of the dust grains would set a lower limit of their distance from the nucleus of 
$R_{\rm min} \ge  0.4pc \times L_{45}^{1/2}T_{1500}^{-2.6} = 0.76~pc$, where the bolometric luminosity estimate above
is assumed,  and the dust sublimation temperature is taken to be 1500 K  \citep{elitzur2006}. 
Unless the absorbing cloud was making its first infall into the central region and destroying its dust
grains in the process, it should have been at a larger distance than this limit when it eclipsed the nucleus.
This corresponds to a maximum orbital velocity of 1100 $\rm km~s^{-1}$, and this upper limit is illustrated
by the vertical limit in Figure 5.

The discrepancy is that the time to travel across 4 $\times$ $10^{15}$ cm
at only a transverse velocity of 1100 $\rm km~s^{-1}$, is $3.6 \times10^7$ seconds, which is
7 times longer than our upper limit to the length of the eclipse event.
Even if the equilibrium dust temperature were 2000 K, the higher inferred
orbital speed would still leave a crossing-time discrepancy of a factor of 5.

Thus it appears that one of our basic assumptions is probably incorrect.  
The short eclipse could have been caused by a rapidly
infalling dust cloud  that approached {\it much closer} than a fraction of
a parsec, at a much higher speed.  This close approach probably 
resulted in the evaporation of the dust, and that could have ended the
quasar reddening that it temporarily produced.

\subsection{Extinction Curve}

We continue exploring our hypothesis that this quasar experienced an eclipse event by a dusty
cloud moving into and out of the line-of-sight to the central engine, all
within 65 days. 
Assuming the intrinsic quasar continuum did not vary, we can compare
its shape before and during eclipse to infer the extinction law produced by
the transiting absorber.
In Figure 6, the gray line shows the extinction curve derived by dividing the green spectrum by the black one in Figure 1. 
It is much steeper than the average  Small Magellanic Cloud (SMC) bar extinction curve \citep{gordon2003} at $\lambda^{-1}>2.5$. 
In recent years, although "grey " extinction curves are often claimed by many investigations \citep{czerny2004,gaskell2004,gaskell2007}, 
a growing number of heavily reddened quasars are detected \citep{wang2005,zafar2012,jiang2013}. 
We attempt to match our steep extinction curve with an SMC-like curve \citep{fitz1990}, 
since SMC-like curves have a steeper UV rise than other observed extinction curves like the Milky Way,  Large Magellanic cloud, 
supernova \citep{FM2007,gordon2003,gallerani2010}. However, our extinction curve is too steep, we could only obtain a poor fit with the interstellar reddening model (red line) in Figure 6. 
The shape of the extinction curve depends on the grain size distribution, as  grains tend to absorb and scatter light most effectively at wavelengths 
comparable to their size. So one possible explanation of the anomalous extinction curve might be 
that large dust grains in the circumnuclear region of a quasar may be destroyed by high energy UV photons or strong shocks 
\citep{jones1994, jones2004}.  
It is widely believed that dust grains form in stellar ejecta (SN and AGB \cite{dwek1998}), while \cite{elvis2002} argued that quasars 
can also form dust grains in the outflowing gas.  Either of these possibilities might lead to an anomalous dust grain distribution.

\subsection{Variable Broad Absorption Lines}
Another alternative is that the extinction was not caused by dust grains,
but rather by variable  blended absorption troughs produced by
atomic gas moving much faster, much closer to the
black hole.  The simplest version of this scenario was that the
same kind of rapid outflow that is seen in front of Broad Absorption Line
quasars, temporarily passed through our line-of-sight to the center.   
This explanation seems reasonable, if there were enough different absorption lines
spread over a sufficiently wide velocity range \citep{hall02,trump06,meusinger2016}.

\acknowledgements
\begin{acknowledgements}
We especially thank the anonymous referee for his/her thorough report and helpful comments
and suggestions that have significantly improved the paper. This work is supported by China Scholarship Council Graduate Fellowship, by the National Science Foundation of China (grants 11473054, 11373056, and U1531245) and by the Science and Technology Commission of Shanghai Municipality (14ZR1447100). This work makes extensive use of SDSS-I/II and SDSS-III data. The SDSS Web Site is http://www.sdss.org/.
\end{acknowledgements}

\begin{figure}[htpb]
\figurenum{1}
\epsscale{1.2}
\plotone{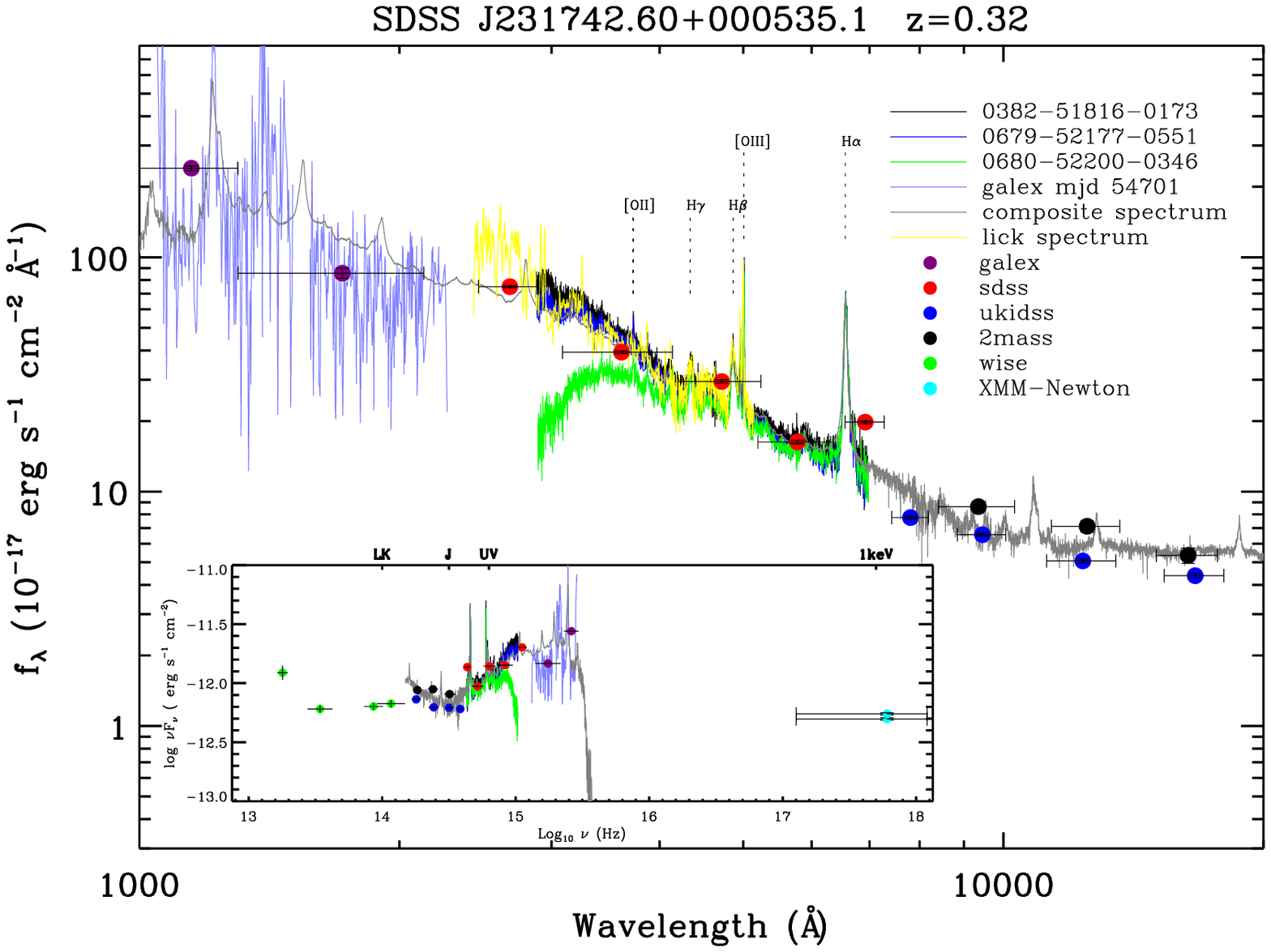}
\caption{\footnotesize The SED of J2317+0005 at rest wavelengths from 1000{\AA} to 20000{\AA}. 
Different spectroscopic and photometric data are presented in various colors. 
The gray spectrum is the quasar composite spectrum \citep{vanden2001,glikman2006}  
A boxcar of 10{\AA} is used to smooth the GALEX (blue line) and Lick spectra (yellow line). 
The other spectra (black, purple and green lines) are SDSS spectra. 
The purple, red, blue, black, green and cyan filled circles are observed by GALEX, SDSS, UKIDSS, 2MASS, WISE, and XMM-Newton, respectively. 
All the fluxes are corrected the Galactic extinction without any normalization. Inset: Data plotted in $\nu$ $\sim$ $\nu F_{\rm \nu}$ diagram. }
\end{figure}

\begin{figure}[htpb]
\figurenum{2}
\epsscale{1.2}
\plotone{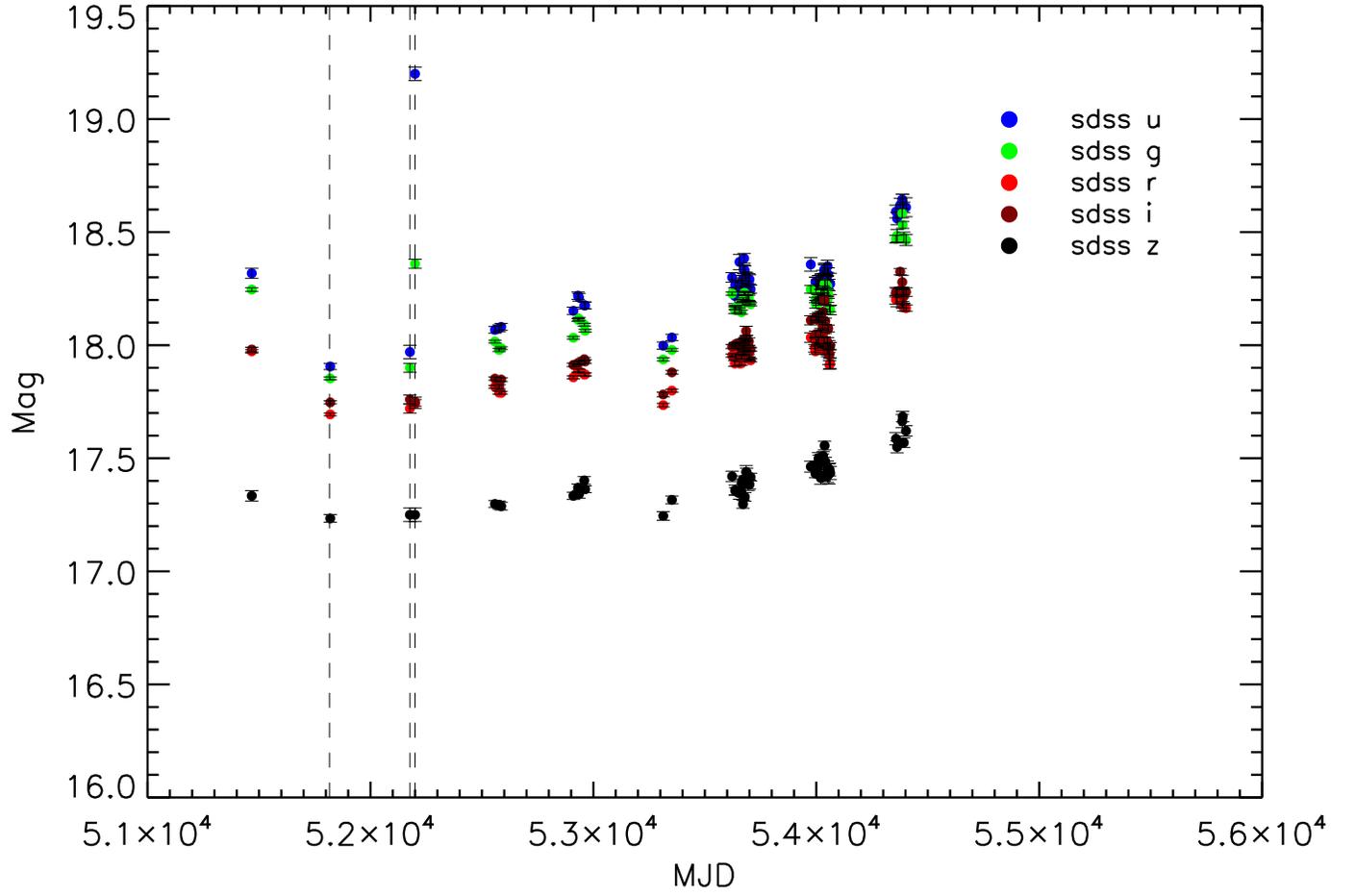}
\caption{\footnotesize Eight-year light curve of J2317+0005 in five bands. 
Three groups of photometric data marked by dashed lines are calculated from $\it u,g,r,i,z$ five bands of spectra (51816, 52177, 52200).}
\end{figure}

\begin{figure}[htpb]
\figurenum{3}
\epsscale{1.2}

\plotone{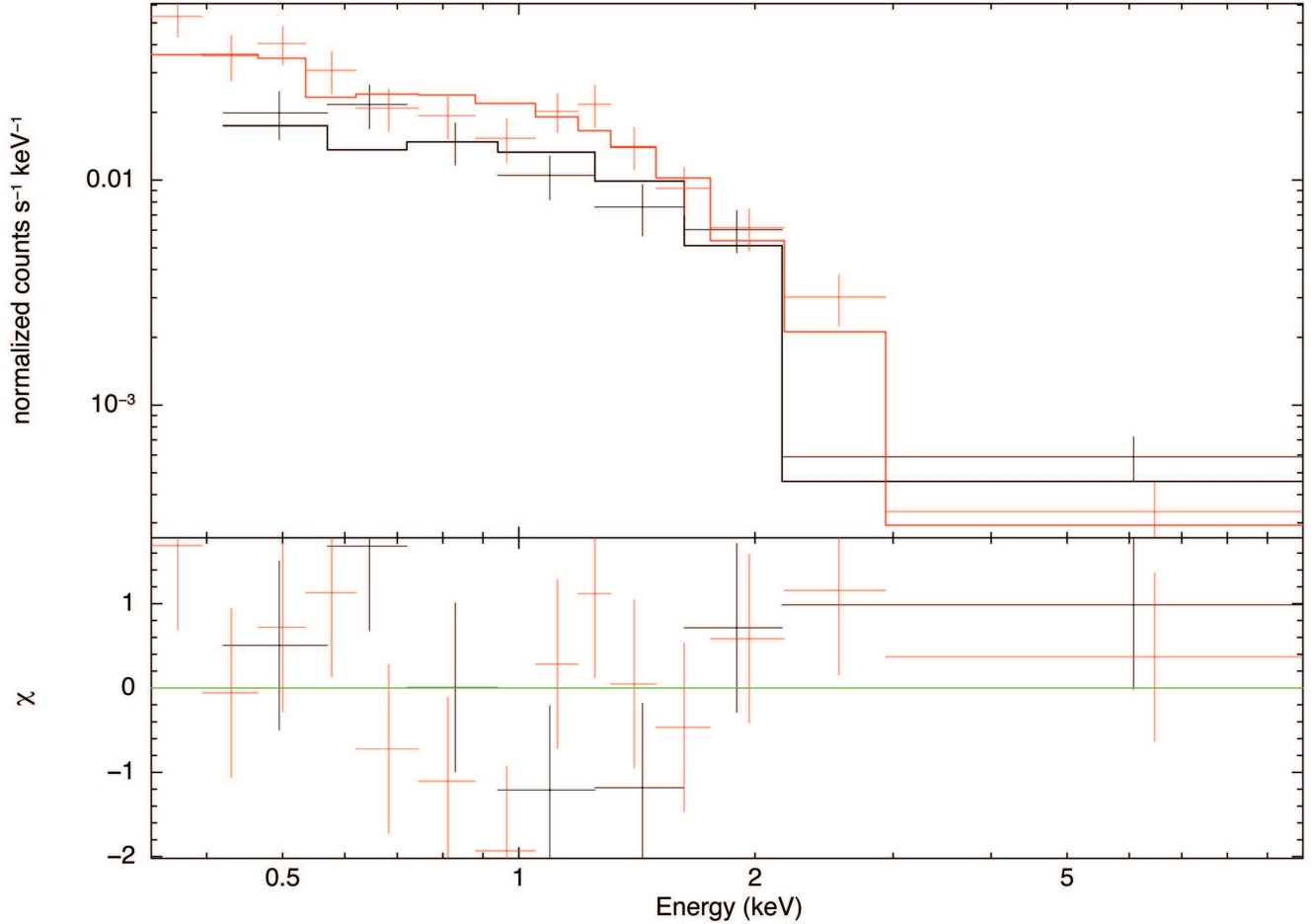}

\caption{\footnotesize X-ray spectra (crosses) fitted by absorption power-law model (line) in the observed frame. 
Possible absorption at the quasar redshift in two observations on 2001 Jun 3 (52063, red, MOS1) and Nov 28 (52242, black, MOS1) 
can be ignored compared to the Galactic absorption.  This indicates the line-of-sight  to central engine is clear on those dates.}
\end{figure}

\begin{figure}[htpb]
\figurenum{4}
\epsscale{1.2}
\plotone{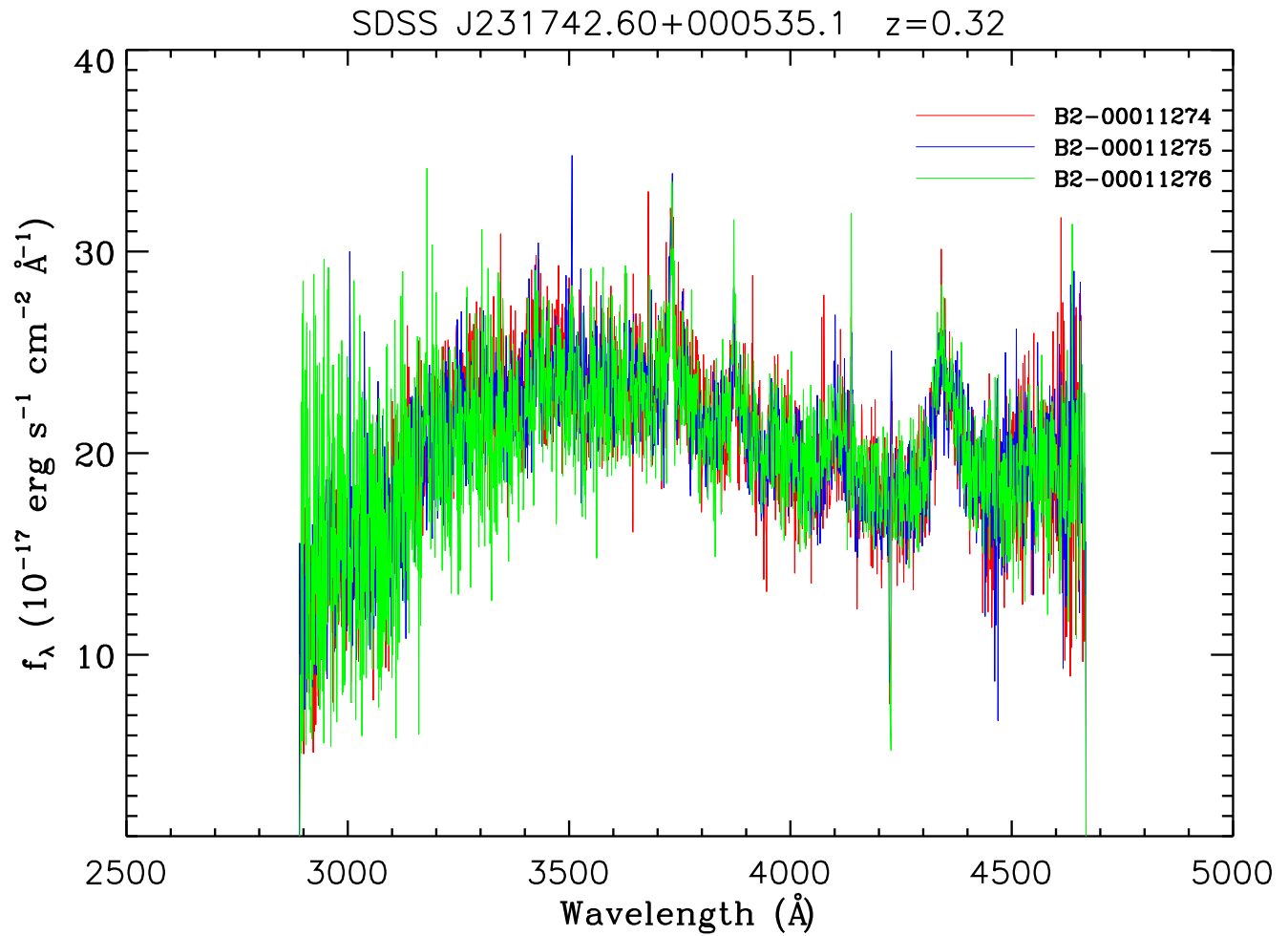}
\caption{\footnotesize Three original blue spectra in observed frame.  The blueward spectrum  of J2317+0005 (52200) is coadded from three individual epochs observed in the same night. }
\end{figure}

\begin{figure}[htpb]
\figurenum{5}
\epsscale{1.2}
\plotone{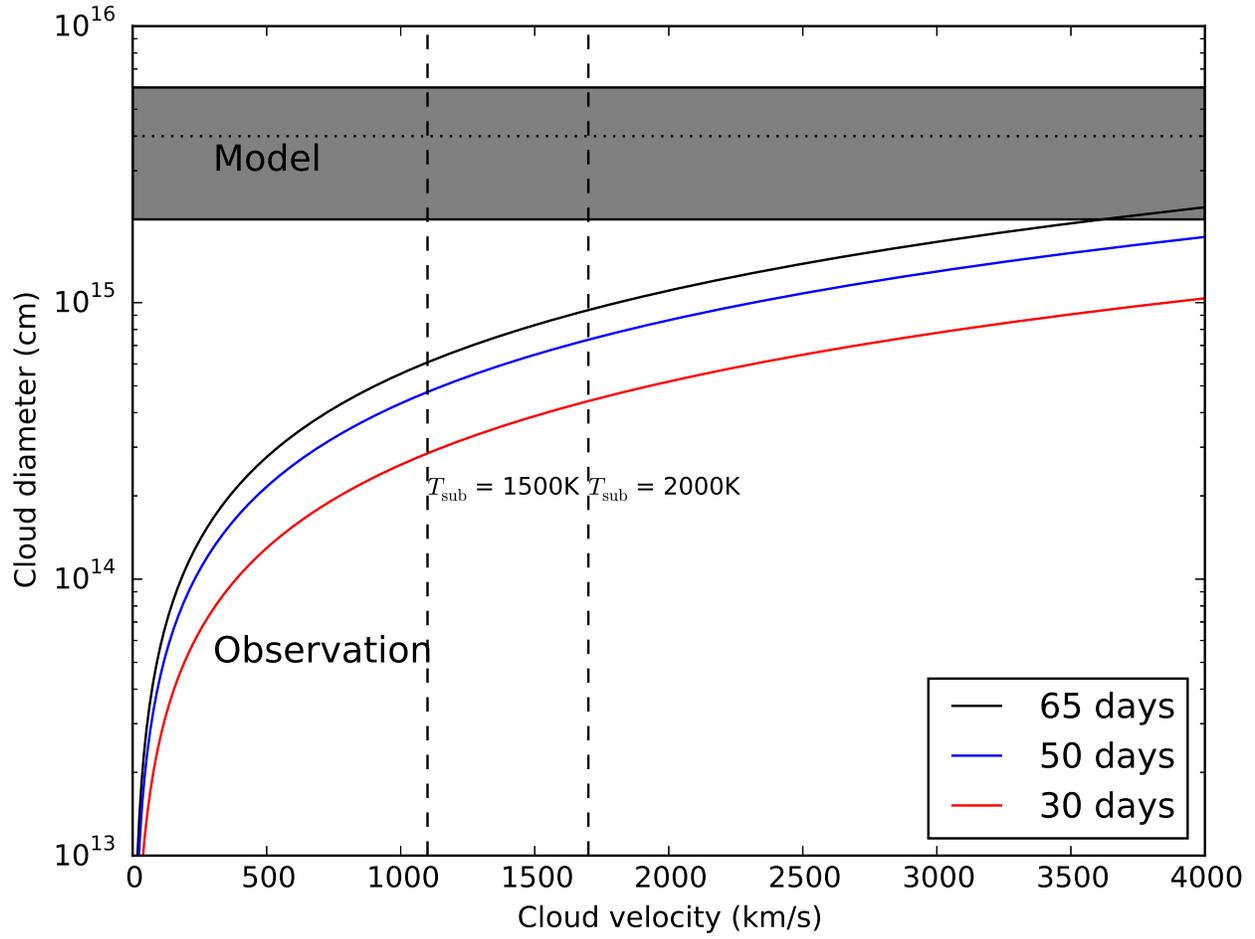}
\caption{\footnotesize Cloud velocity versus diameter. The solid black, blue and red lines are calculated with the eclipse period of 64, 50, 30 days, respectively. Two vertical dashed lines are 
the upper limits of velocity at 1500 and 2000K dust sublimation temperature. The horizontal dotted line is the estimated 3000\AA\ continuum-emitting region with $\pm$ 50\%\ uncertainty (grey band). The cloud size estimated from the observation is a little lower than emitting region predicted by accretion disk model.}
\end{figure}

\begin{figure}[htpb]
\figurenum{6}
\epsscale{1.2}
\plotone{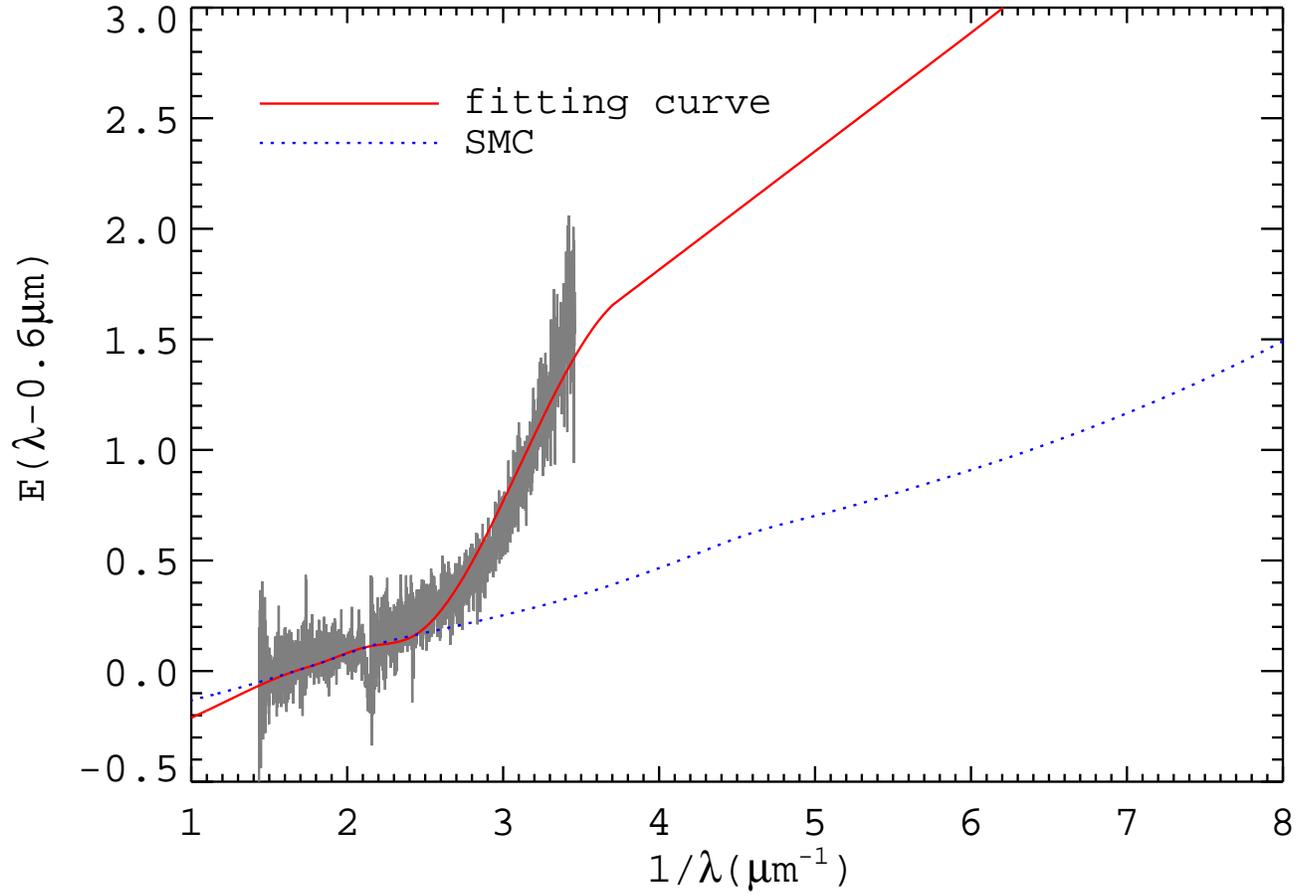}
\caption{\footnotesize Extinction curve caused by clumpy obscuration. The inferred extinction curve (52200) of J2317+0005 is much steeper 
than the average SMC Bar extinction (blue dotted line) from \cite{gordon2003}.  The red line shows the best-fit model from \cite{fitz1990}.}
\end{figure}

\begin{table}\large
\begin{center}
\caption[]{total flux of emission lines ($10^{-17} erg~cm^{-2}~s^{-1}$)}
\label{table3}\small
\begin{tabular}{ccccccc}
\tableline
\tableline

MJD & 51816 &52177 &52200 &57282\\
\hline
[OII]3726,3729  & 170.4$\pm$33 & 186.0$\pm$34 &151.3$\pm$30& ...\\
H${\gamma}$& 167.7$\pm$38 & 165.9$\pm$33& 159.9$\pm$33&...\\
H${\beta} $&1164.0$\pm$98 & 1194.0$\pm$110 & 1133.0$\pm$107& 1150.7$\pm$111&  \\
$\rm [OIII]5007$   &1005.0$\pm$88 & 939.0$\pm$93& 914.2$\pm$90& 920.4$\pm$101&\\
H${\alpha}$+[NII] &4699.3$\pm$412& 4657.1$\pm$392& 4751.9$\pm$421& 4784.1$\pm$380\\
\hline

\tableline
\end{tabular}
\end{center}
\end{table}

\begin{table}
\renewcommand{\arraystretch}{1.2}
\begin{center}
\caption[]{Photometric Data in AB mag}
\label{table1}\small
\begin{tabular}{ccccc}
\tableline
\tableline

Band &   Value&     Facility   &    Date   &MJD   \\
 &  (AB mag)    &   &                                                                 (UT)&       \\
\hline
FUV  &  19.14 $\pm$ 0.07&  GALEX  & 2003 Aug 25 &  52876  \\  
NUV  &  18.91 $\pm$ 0.04 & GALEX  & 2003 Aug 25 &  52876\\
$\it u$ & 17.69 $\pm $ 0.01 & SDSS & 2001 Oct  2 & 51819 \\
\it g & 17.85  $\pm$ 0.01 & SDSS & 2000 Oct  2 & 51819 \\
\it r&17.69 $\pm$ 0.01 & SDSS &2000 Oct  2 & 51819 \\
\it i&17.74  $\pm$ 0.01 & SDSS &2000 Oct  2 & 51819 \\
\it z&17.23 $\pm$ 0.02 & SDSS &2000 Oct 2 & 51819\\
\it J &17.31 $\pm$  0.11& 2MASS&2000 Aug 25 &51781\\
\it H&16.89 $\pm$ 0.10& 2MASS & 2000 Aug 25 & 51781\\
$\it K_{s}$ &16.61 $\pm$ 0.10& 2MASS & 2000 Aug 25 & 51781\\
\it Y&17.80 $\pm$ 0.02 & UKIDSS & 2006 Nov 21 & 54060 \\ 
\it J& 17.59 $\pm$ 0.02 & UKIDSS & 2006 Nov 23 & 54062 \\
\it H& 17.28 $\pm$ 0.02 & UKIDSS & 2005 Oct 26 & 53669 \\
$\it K_{s}$& 16.77 $\pm$ 0.01&UKIDSS & 2005 Oct 26 & 53669 \\
W1&16.40$\pm$ 0.03 & WISE & 2010 Jun 10 & 55357 \\	
W2&16.13 $\pm$ 0.03 & WISE & 2010 Jun 10 & 55357 \\	
W3&15.27 $\pm$ 0.05 & WISE & 2010 Jun 10 & 55357 \\	
W4&13.70 $\pm$ 0.09 & WISE & 2010 Jun 10 & 55357 \\
\hline

\tableline
\end{tabular}
\end{center}
\end{table}

\begin{table}\tiny
\begin{center}
\caption[]{Spectroscopic Data}

\label{table2}
\begin{tabular}{ccccccc}
\tableline
\tableline
Range &   slit&     $\lambda/\delta\lambda$ &    Exp. Time  &  Instrument &  Date   &MJD   \\
({\AA}) &  (arcsec)    &   &                                     (s)&                            & (UT)       \\
\hline
1344-1786&  slitless& 200     &1460&      GALEX/Grism  & 2009 Mar 25 &  54915\\ 
1771-2831&    slitless&   90       &1460&      GALEX/Grism  &2009 Mar 25 &   54915\\
3100-6200&    2.0&  3390   & 1800&  Lick/Shane/KAST& 2015 Sep 18&   57282\\ 
3800-7310&    2.0&  4748   & 1800&  Lick/Shane/KAST& 2015 Sep 18&   57282\\ 
3800-9200&   3.0&   2000&   3600&  SDSS&  2000 Sep 29& 51816\\
3800-9200&   3.0&   2000&   4203&  SDSS&  2001 Sep 25& 52177\\
3800-9200&   3.0&   2000&   3904&  SDSS&  2001 Oct 18& 52200\\ 
   
\hline

\tableline
\end{tabular}
\end{center}
\end{table}

\end{CJK}
\end{document}